\documentclass[%
 reprint,
showpacs,
 amsmath,amssymb,
 aps,
prl,
]{revtex4-1}

\usepackage[pdftex]{graphicx}
\usepackage{dcolumn}
\usepackage{bm}

\def\be{\begin{equation}}
\def\ee{\end{equation}}
\def\ba{\begin{eqnarray}}
\def\ea{\end{eqnarray}}

\def\rd{\mathrm{d}}
\def\rD{{\rm D}}
\def\CL{{\cal L}}
\def\g{\mathfrak{g}}

\newcommand*{\R}{{\mathbb R}}

\begin{document}

\title{Curving Yang-Mills-Higgs Gauge Theories}

\author{Alexei Kotov}\email{oleksii.kotov@uit.no}
\affiliation{Departamento de Matem\'{a}tica, Universidade Federal do Paran\'{a} \\ C.P. 019081, 81531-990
Curitiba - Paran\'{a}, Brazil}

\author{Thomas Strobl}\email{strobl@math.univ-lyon1.fr}
\affiliation{Institut Camille Jordan,
Universit\'e Claude Bernard Lyon 1 \\
43 boulevard du 11 novembre 1918, 69622 Villeurbanne cedex,
France
}%
\date{July 14, 2015}
\begin{abstract}
We present a Yang-Mills-Higgs (YMH) gauge theory in which structure constants of the gauge group may depend on Higgs fields. The data of the theory are encoded in the bundle $E \to M$, where the base $M$ is the target space of Higgs fields and fibers carry information on the gauge group. $M$ is equipped with a metric $g$ and $E$ carries a connection $\nabla$. If $\nabla$ is flat, $R_\nabla=0$, there is a local field redefinition which gives back the standard YMH gauge theory. If $R_\nabla \neq 0$, one obtains a new class of gauge theories.  In this case, contrary to the standard wisdom of the YMH  theory, the space $(M, g)$ may have no isometries. We build a simple example which illustrates this statement.
\end{abstract}

\maketitle

\section{Introduction}
One way of introducing General Relativity is to start from Special Relativity: by rewriting equations such as the one for the free movement of a particle in a coordinate independent manner,  one notices that this introduces Christoffel symbols $\Gamma^i_{jk}(x)$ corresponding to a connection $\nabla$ that is flat, $R_\nabla = 0$. Dropping the latter condition on $\nabla$ is an essential step in the development of Einstein's theory of gravity, where, in general,  $R_\nabla \neq 0$.

We will use this strategy in the context of the coupled Yang-Mills-Higgs (YMH) theory, 
i.e.~the framework used as the bosonic sector of the current Standard Model of particle physics. We will make the theory covariant with respect to Higgs-dependent basis changes of the structural Lie algebra. As in the previous paragraph, there will be a connection $\nabla$: if flat, the theory is equivalent to an ordinary YMH-theory, otherwise it is of new type.

Let us first collect the data used for conventional YMH theories. We restrict to trivial bundles for simplicity. The gauge fields or YM-connections are Lie algebra valued 1-forms $A=A^a \otimes e_a \in \Omega^1(\Sigma,\g)$ on a 
$d$-dimensional Lorentzian spacetime manifold $\Sigma$. Their kinetic term is given by the square of the YM-curvature $F^a = \rd A^a + \frac{1}{2} C^a_{bc} A^b \wedge A^c$,
\begin{equation}\label{YM}
S_{YM}[A]  =  \int\limits_\Sigma \frac{1}{2} \kappa_{ab} F^a \wedge *F^b \, .
\end{equation}
Here $\kappa$ is an ad-invariant metric  on $\g$, $*$ denotes the Hodge-duality operator induced by the metric on 
$\Sigma$, and $C^a_{bc}$ are the structure constants, $[e_a,e_b] = C^c_{ab} e_c$.

The Higgs fields correspond to a map $X \colon \Sigma \to M$, where $M$ is an n-dimensional Riemannian manifold whose metric $g$ is invariant with respect to $\g$: if $e_a$ is represented by $\rho_a \in \Gamma(TM)$, then 
$\CL_{\rho_a}{}g=0$. Choosing local coordinates $(x^i)_{i=1}^n$ on $M$, their pullback by $X$ yields 
$n$ scalar fields $X^i=X^*(x^i)\in{}C^\infty(\Sigma)$. 
In addition to these data, conventionally one also has a $\g$-invariant function $V$ on $M$, giving rise to the Higgs potential. 

The coupled YMH functional then has the form $S_{YMH} [A,X] = S_{YM}[A] + S_{Higgs}[X,A]$ where
\begin{eqnarray} \label{Higgs}
S_{Higgs} [X,A] & = &  \int\limits_\Sigma  
\frac{1}{2} g_{ij}(X) \, \rD X^i \wedge * \rD X^j  + 
*V(X) \, .
\end{eqnarray}
The interaction of $A$ and $X$ is effectuated by means of the covariant derivatives (minimal coupling)
\begin{equation}\label{DX}
\rD X^i = \rd X^i - \rho_a^i(X) A^a \, .
\end{equation}
Clearly, a
Higgs-dependent change of basis of the gauge fields
$A^a \to (M^{-1})^a_b(X) A^b$ leaves the \emph{form} of \eqref{Higgs} and \eqref{DX} unchanged, while \eqref{YM} changes rather drastically. There not only the new  $\kappa$-coefficients become Higgs-field dependent, but, most importantly, also the structure constants $C^a_{bc}$, thus turning them into \emph{structure functions}. Such a redefinition of fields gives rise to a connection
\begin{equation}\label{omega}
\omega^a_b = (M^{-1})^a_c \rd M^c_b \, .
\end{equation}
The flatness of these connection coefficients shows that the apparent modifications of the coupled theory in terms of $S_{YM}$ can be made undone again, restoring the original form \eqref{YM} and \eqref{Higgs} 
of the theory. Our interest will lie in cases, where $\omega$ is not necessarily flat. 

\section{Covariantizing and Curving the graded target space}
An $X$-dependent change of $A^a$ corresponds to an inverse change of the basis $e_a$, leaving $A = A^a \otimes e_a$ unchanged. This is understood best by introducing the vector bundle $E=M\times \g \to M$, turning the problem into one of finite dimensional geometry on the target space of the theory. Any Lie algebra element can be viewed as a section of $E$ now and 
\begin{equation}\label{change}
e_a \to M_a^b(x) e_b
\end{equation}
corresponds to a change of the basis in $E$. Note that $E$ carries a natural flat connection. If in the original basis $\nabla e_a = 0$, then after the change \eqref{change}, the connection coefficients become precisely as in Eq.~\eqref{omega}. Here we used that in a general local frame $(e_a)_{a=1}^r$ of $E$, a connection $\nabla \colon \Gamma(E) \to \Gamma(T^*M \otimes E)$ satisfies
\begin{equation}\label{conn}
\nabla e_a = \omega_a^b \otimes e_b \, .
\end{equation}

We will need the transformation property of $C^a_{bc}$ under a change of basis. This can be deduced most easily from the representation property of the fundamental vector fields $\rho_a$, $[\rho_a,\rho_b] = C^c_{ab} \rho_c$, yielding:\footnote{Straight brackets $[\ldots ]$ indicate antisymmetrization, round ones $( \ldots )$ symmetrization. Commas denote derivatives with respect to $x$: $f_{,i} \equiv \partial_i f\equiv \frac{\partial f}{\partial  x^i}$.} \begin{equation}\label{Cchange}
C^a_{bc} \to (M^{-1})^a_d M^e_b M^f_c  C^d_{ef} - 2(M^{-1})^a_d M^e_{[b} M^d_{c],i}  \rho_e^i\, , 
\end{equation} 
where on the r.h.s.~we recognize that the inhomogeneous term is proportional to the connection coefficients \eqref{omega}. We observe that the structure constants
turned into structure functions $C^a_{bc}(x)$. While  we will drop the condition on $\nabla$ to be flat, we need to determine what replaces the Jacobi identity in the case of such structure functions. 

For this purpose, we take recourse to a BRST-model of the $\g$-action on $M$: Consider 
the infinitesimal transformations $\delta x^i := \beta^a \rho^i_a(x)$ for some parameters $\beta^a$. The idea of BRST is to declare $\beta^a$s to be odd, i.e.~anticommuting, parameters and to make them also transform such that the transformations square to zero. This corresponds to the nilpotency of the following odd vector field 
\begin{equation}\label{Q}
Q = \beta^a \rho_a^i \frac{\partial}{\partial x^i} - \frac{1}{2} C^a_{bc} \beta^b \beta^c \frac{\partial}{\partial \beta^a}  \, , 
\end{equation}
which is of the standard form of the BRST charge known from Yang-Mills gauge theories \cite{Henneaux-Teitelboim}, replacing the field space by a finite dimensional toy model with coordinates $x^i$ and $\beta^a$ of degrees zero and one, respectively. These, in turn, can be viewed as coordinates on $E$---or better on $E[1]$, indicating by the notation that the fiber-linear coordinates $\beta^a$ have been declared to have degree one. Corresponding to sections in $E^*$, they transform according to $\beta^a \to (M^{-1})^a_b \beta^b$ with respect to \eqref{change}. This is also compatible with \eqref{Cchange}. 

$Q$ is a vector field on $E[1]$ squaring to zero: $Q^2=0$. This is a coordinate independent feature, insensitive to \eqref{change} in particular. In fact, any graded manifold described locally by degree 0 and degree 1 coordinates takes the form of a shifted vector bundle $E[1]$ and any degree +1 vector field evidently has the form of \eqref{Q} for some coefficient functions $\rho_a^i(x)$ and $C^a_{bc}(x)$. A vector bundle $E$ equipped with a nilpotent vector field \eqref{Q} on $E[1]$  is called a Lie algebroid (cf.~\cite{Mackenzie,daSilva-Weinstein} on the mathematics of Lie algebroids and \cite{Vaintrob,Gruetzmann-Strobl} for its super-geometrical formulation). 

{}From the present perspective, the transition from abelian Lie algebras, which \emph{are} just vector spaces, to non-abelian ones together with their action on a manifold $M$ is more drastic than the one from Lie algebras to Lie algebroids: The first step can be viewed as introducing the BRST operator \eqref{Q}, while the second step permits besides $\rho^i_a$ also $C^a_{bc}$ to become $x$-dependent. In both cases $Q^2=0$ contains all the axioms of the respective notion. 

The connection \eqref{conn} and the Lie algebroid structure \eqref{Q} have to satisfy a compatibility condition that will be imposed on us by gauge invariance below. To formulate it, we first observe that the structure constants/functions $C^a_{bc}$ do not behave like tensors according to \eqref{Cchange}. This can be cured by means of the connection coefficients \eqref{conn}: $t^a_{bc} := C^a_{bc} - 2\rho_{[b}^i \omega^a_{c]i}$ behaves like a tensor w.r.t.~\eqref{change}, i.e.~$t \in \Gamma(E \otimes \Lambda^2 E^*)$. The compatibility condition is\footnote{This condition appeared in  \cite{Mayer-Strobl} as $S=0$ already. For a geometric interpretation cf.~\cite{Killing}.}
\begin{equation}\label{S}
(\nabla t)^a_{bci} = 2\rho_{[b}^j (R_\nabla)^a_{c]ji} \, .
\end{equation}
We observe that the curving of the Yang-Mills-Higgs theory is \emph{completely} governed by the curvature $R_\nabla$ of the connection \eqref{conn}: Indeed, if $R_\nabla =0$, one may choose a locally constant frame $e_a$ such that $t$ equals to the structure functions, and 
then \eqref{S} enforces those to be  \emph{constants}. Thus, locally, and that is all what we are interested in here, $R_\nabla =0$ implies $E \cong U \times \g$ for some Lie algebra $\g$. 

The fields $X$ and $A$ together can be viewed as arising from a degree preserving map $a \colon T[1]\Sigma \to E[1]$.  $T[1]\Sigma\equiv T\Sigma[1]$  indicates that fiber-linear functions on $T\Sigma$, i.e.~sections in $\Gamma(T^*\Sigma) \cong \Omega^1(\Sigma)$, are considered to be of degree one; so, general functions on 
$T[1]\Sigma$ are just differential forms on $\Sigma$. Then $a^*(x^i) = X^i \in \Omega^0(\Sigma)$ and $a^*(\beta^a) = A^a \in \Omega^1(\Sigma)$ reproduce the previous fields.
Thus, $S_{YMH}$ and its curved generalization
are a kind of ``super''-sigma model, where the target is a Q-manifold $(E[1],Q)$ corresponding to a Lie algebroid $E \to M$. It can  deviate from $E=M\times\g$ only if the curvature 
$R_\nabla$ of the connection \eqref{conn} does not vanish.

\section{Gauge Transformations and \\Field Strength for the Gauge fields}
While the gauge transformations of the Higgs fields are already covariant, $\delta X^i = \varepsilon^a \rho_a^i(X)$ for any $\varepsilon = \varepsilon^a e_a \in \Gamma(X^*E)$, the ones for the gauge fields take the form 
$\delta A^a = \rd\varepsilon^a + C^a_{bc}A^b \varepsilon^c$ only in a locally flat basis. With \eqref{Cchange} and using \eqref{omega}, this turns into
\begin{equation}\label{Atrafo}
\delta A^a = \rd \varepsilon^a + C^a_{bc}(X) A^b \varepsilon^c + \omega^a_{bi} \varepsilon^b \rD X^i \, .
\end{equation} 
We now postulate this formula also for a non-flat connection \eqref{conn}, while for $R_\nabla = 0$ and a flat frame it evidently reduces to the standard formula.  Eq.~\eqref{Atrafo} was found already in \cite{BKS} and \cite{withoutsymmetry} by other arguments. The transformations \eqref{Atrafo} close off-shell,  \emph{iff} \eqref{S} holds true \cite{Mayer-Strobl}.

Next we turn to the replacement of the Yang-Mills part \eqref{YM}, necessary for the curving of the theory. Again applying our ``covariantization strategy'', one finds that the Yang-Mills curvature $F^a$ becomes a tensor with respect to \eqref{change}: While $A \in \Gamma(X^*E \otimes T^*\Sigma)$, its field strength $F \in \Gamma(X^*E \otimes \Lambda^2 T^*\Sigma)$ takes the form: 
\begin{eqnarray}
F^a &=& (\rD A)^a + \frac{1}{2}  t^a_{bc}(X) A^b \wedge A^c \label{F} \\
(\rD A)^a &\equiv & \rd A^a + \omega^a_{bi}(X) \, \rD X^i\wedge A^b \, . \nonumber
\end{eqnarray}
In standard YM theory, $A$ and $F$ have the mathematical meaning of a connection and its curvature. Here, we call $A$ simply a gauge field and $F$ its field strength (cf.~also \cite{KS07,Gruetzmann-Strobl})---or ``YM-connection'' and ``YM-curvature'', respectively, so as to distinguish them well from the connection $\nabla$ and its curvature $R_\nabla$: together with the target Lie algebroid, $\nabla$ is \emph{fixed} for a given ``curved YMH theory''---like the Lie algebra $\g$ in the standard situation. 

It remains to calculate the behavior of \eqref{F} with respect to the gauge transformations \eqref{Atrafo}. This calculation was performed already in \cite{Mayer-Strobl}: Using Eq.~\eqref{S} and the structural identities following from $Q^2=0$ with \eqref{Q}, one obtains 
$\delta F^a =  (C^a_{bc} - \omega^a_{ci} \rho_b^i) \varepsilon^c F^b + \frac{1}{2} R^a_{bij}\varepsilon^b \rD X^i \wedge \rD X^j \, .$
This equation shows that the field strength defined in Eq.~\eqref{F} transforms into itself if and only if $R_\nabla = 0$. 

At this point one might think, as in \cite{Mayer-Strobl}, that the present attempt for a curved generalization fails. However, there is a way out: Let us consider the following general ansatz  starting with $\rd A$ and followed by a quadratic expression in $A$ and $\rd X$ or $\rD X$: 
$$
G^a = \rd A^a + \frac{1}{2} {\bar C}^a_{bc} A^b \wedge A^c + {\bar \omega}^a_{bi}  \rD X^i \wedge A^b + \frac{1}{2} B^a_{ij} \rD X^i \wedge \rD X^j  
$$
where the coefficients ${\bar C}^a_{bc}$, ${\bar \omega}^a_{bi}$, and $B^a_{ij}$ are at this point arbitrary functions of $X$. We now require that $G^a$ transforms into itself. Using the previous formulas for the gauge transformations
which also imply
\begin{equation}\label{deltaDX}
\delta \left(\rD X^i\right)=   \varepsilon^a \left(\rho_{a,j}^i - \rho_b^i \omega_{aj}^b \right) \rD X^j
\, ,
\end{equation}
a straightforward calculation 
yields that necessarily ${\bar C}^a_{bc}=C^a_{bc}$, ${\bar \omega}^a_{bi} = \omega^a_{bi}$, and that
\begin{equation}\label{curvature}
 (R_\nabla)^a_b + \CL_{\rho_b} B^a - \omega^c_{b}\wedge \iota_{\rho_c} B^a +\iota_{\rho_b}(\omega^a_c) \, B^c + t^a_{bc} B^c =0\, ,
\end{equation}
where $B^a = \frac{1}{2} B^a_{ij}(x) \rd x^i \wedge \rd x^j$ and $B = B^a \otimes e_a \in \Gamma(\Lambda^2 T^*M \otimes E)$. The second, third, and fourth term combine into a ``covariantized Lie derivative'': $\left([\rD,\iota_\rho]B\right)^a_b$. With this choice, 
\begin{eqnarray}
G^a &=& F^a + \frac{1}{2}B^a_{ij}(X) \rD X^i \wedge \rD X^j \, , \label{G}\\
\delta G^a &=&  (C^a_{bc} - \omega^a_{ci} \rho_b^i) \varepsilon^c G^b \, .\label{deltaG}
\end{eqnarray} 
A gauge invariant action functional can be formed  by ``squaring'' the quantity $G$.
The additional contribution 
to $F$ in \eqref{G}, governed by an $E$-valued 2-form $B$ on the Higgs target manifold $M$, is essential: there is no \emph{non-trivial} deformation of the YMH setting without it.

\section{The gauge invariant, curved action}
Now we can present the gauge invariant curved action functional. It takes the form
\begin{equation}\label{CYMH}
S_{CYMH}[A,X]  =  \int\limits_\Sigma \frac{1}{2} \kappa_{ab}(X) G^a \wedge *G^b  + S_{Higgs}[X,A]\, .
\end{equation}
Gauge invariance of $S_{CYMH}$ requires the metric $g$ on $M$, entering $S_{Higgs}$ as in \eqref{Higgs}, as well as the fiber metric $\kappa \in \Gamma(S^2 E^*)$ to satisfy appropriate conditions. 

To obtain the one for $g$, we first covariantize $\CL_{\rho_a}g=0$ with respect to \eqref{change}. Using the replacement \eqref{omega}, this yields
\begin{equation}\label{KillingLie}
\left(\CL_{\rho_a} g\right)_{ij}= 2\omega_{a(i}^b \rho_{j)b} \, ,
\end{equation}
where on the r.h.s.~the upper index of $\rho$ was lowered by means of the metric. This equation can be also obtained directly from \eqref{deltaDX} for arbitrary $\omega$-coefficients. Eq.~\eqref{KillingLie} can be rewritten as $\rho_{a(i;j)}=0$, where the semicolon denotes a covariant differentiation---with respect to \emph{both} of the preceding tensor indices, using $\nabla$ for the first one and the Levi-Civita connection of $g$ for the second one.

We see that if $\nabla$ is not flat, the metric $g$ entering the kinetic term of the Higgs field does no more need to have an isometry so as to ensure gauge invariance of the curved theory (while for $R_\nabla = 0$, in a constant frame, the $\rho_a$s are local Killing vectors). Eq.~\eqref{KillingLie} was found already in \cite{withoutsymmetry}; the present paper can be viewed as an extension of the previous one by adding an additional kinetic term for the gauge fields. Geometrically Eq.~\eqref{KillingLie} implies that the orbits in $M$ generated by $\rho_a$ form a Riemannian foliation; we refer to \cite{Killing} for this and further geometrical features. 

The transformation property required for $\kappa$ follows directly from \eqref{CYMH} and \eqref{deltaG}. It can be formulated best in terms of what is called an $E$-connection\footnote{Cf.~\cite{Mayer-Strobl} for further details on the following.}: The Q-structure \eqref{Q} equips $\Gamma(E)\ni s,\tilde{s}$ with a Lie bracket $[s,\tilde{s}] = \left(s^a \tilde{s}^b C^c_{ab} +  \rho(s)\tilde{s}^c - \rho(\tilde{s})s^c\right) e_c$ where $\rho(s)\equiv s^a \rho_a$. Using $\nabla$, this induces ${}^E\! \nabla_s \tilde{s} := [s,\tilde{s}] - \nabla_{\rho(\tilde{s})}s$, which permits us to perform  covariant derivatives of $E$-tensors along sections of $E$. The condition on $\kappa$ is just that it should be $E$-covariantly constant:
$$
{}^E\! \nabla \: \kappa  = 0 \, .
$$
${}^E\! \nabla$, defined canonically in any Lie algebroid $E$ equipped with a connection $\nabla$, is a generalization of the adjoint representation. Indeed, as a consequence of \eqref{S}, its $E$-curvature vanishes, which is tantamount to the 
representation property: $[{}^E\! \nabla_{e_a} , {}^E\! \nabla_{e_b}] = C^c_{ab}(x) \, {}^E\! \nabla_{e_c}$. 

Finally, for gauge invariance of \eqref{CYMH}, the Higgs potential has to be invariant, $\rho_a^i \partial_i V = 0$. Note that this condition is invariant under base changes \eqref{change}.

\section{A curved example with Lie algebras}
For a theory of the type \eqref{CYMH} we need to specify: A pseudo-Riemannian manifold $\Sigma$, a Lie algebroid $E \to M$ with a connection $\nabla$, a function $V$ and an $E$-valued 2-form $B$ on $M$, and metrics $g$ and $\kappa$ on $M$ and $E$, respectively. These data have to satisfy several compatibilities. 

To not distract from the essentials, we will provide a curved example in the context of an ordinary Lie algebra, in fact, even an abelian one. This illustrates within a simple setting the qualitatively new things one can do when permitting $R_\nabla \neq 0$, like constructing a gauge theory in the absence of any isometry of the metric $g$. 

Let $\Sigma$ be 4-dimensional Minkowski space and $E$ a trivial real line bundle over $M=\R^2 \ni (x,y)$ with $Q=\beta \partial_y$. Furthermore, $g := (\rd x)^2 + \exp(\lambda xy) (\rd y)^2$. One can show that this metric has \emph{no} local isometries \emph{iff} $\lambda \neq 0$; in particular, obviously $\partial_y$ is not a Killing vector. Still $y \mapsto y + \mathrm{const}$ can be gauged \cite{withoutsymmetry}, since \eqref{KillingLie} is satisfied for $\omega = \frac{\lambda x}{2} \rd y$. $R_\nabla = \rd \omega$ is non-zero for $\lambda \neq 0$; so here $\lambda$ is a  deformation parameter of an otherwise simple abelian YMH theory. It remains to fix $B$, $\kappa$, and $V$: $B = -\frac{\lambda y}{2}\rd x \wedge \rd y$ and $\kappa = \exp(\lambda xy)$
do the job, while $V$ can be an arbitrary function of $x$, e.g.~$V= - \mu x^2 + \nu x^4$. 

Although the action of two scalar fields $X$ and $Y$
$$
S_{0}[X,Y] = \int_{\R^4} \left(\partial_\mu X \partial^\mu X + e^{\lambda XY} \partial_\mu Y \partial^\mu Y + V(X)\right)
\rd^4 \sigma
$$
does not have any global symmetries except for $\lambda = 0$, the replacement $\partial_\mu Y \longrightarrow \rD_\mu Y \equiv \partial_\mu Y - A_\mu$, turning $S_0$ into the form \eqref{Higgs}, has the gauge symmetry 
\begin{equation} \label{gauge}
\delta Y = \varepsilon(\sigma) \quad , \qquad \delta A_\mu = \partial_\mu \varepsilon + \frac{\lambda}{2} \varepsilon X \left( \partial_\mu Y - A_\mu \right) \, .
\end{equation}
The theory $S_{Higgs}[X,Y,A]$ by itself is classically equivalent to a sigma model with the target quotient $\R^2/\R\cong \R\ni x$, thus here  simply to $S[X] =  \int_{\R^4} \left(\partial_\mu X \partial^\mu X + V(X)\right)
\rd^4 \sigma$. In general the quotient $M$ modulo the flow generated by the vector fields $\rho_a$ is not a smooth manifold; then the gauged theory $S_{Higgs}$ provides a smooth, field theoretic resolution or description of this ``sigma model with singular target space'' \cite{withoutsymmetry}.
 
The addition of a kinetic term for the gauge fields,
\begin{eqnarray}
S_{kin} &=& \int_{\R^4} e^{\lambda XY} G_{\mu \nu} G^{\mu \nu}
\rd^4 \sigma \, , \label{Glambda} \\
G_{\mu \nu} &=& 2 \partial_{[\mu} A_{\nu]} 
+ \frac{\lambda}{2}\left(X \partial_{[\mu} Y A_{\nu]}+Y\rD_{[\mu} Y \partial_{\nu]} X \right) \nonumber  , 
\end{eqnarray}
destroys the latter feature, also for $\lambda = 0$. However, it is mandatory from the physical perspective, where a square of $\rd A$ is needed
for the gauge fields to describe the propagation of interaction particles like the photons.

All formulas reduce to the most standard 4d abelian YMH model if $\lambda = 0$. $R_\nabla \propto \lambda$ parametrizes a non-trivial deformation of it, consistent with gauge symmetry. Note that the deformation is analytic in $\lambda$, but not polynomial.

\section{Conclusion and Outlook}
Higgs-field dependent changes of the Lie algebra basis for gauge fields are not new, they were e.g.~considered in the context of soliton solutions or to describe field configurations with symmetries. We went one step further though: We consider also inherently Higgs-field dependent bases. If the  structural curvature $R_\nabla =0$, we know that there exists a frame in which the coupled system takes standard form (at least locally). There can be two reasons for  $R_\nabla \neq 0$ (or both simultaneously): either the Higgs sector has no ordinary type of symmetries, but the generalized invariance discussed in \cite{withoutsymmetry}, or the coupled YMH-system
is governed by a Lie algebroid whose structure functions $C^a_{bc}(x)$ cannot be ``rectified'' by a change of coordinates. 

While in the Lie algebroid Yang-Mills theories constructed in \cite{LAYM} (type I LAYM theory) the scalar fields remain topological, in the present ``Curved YMH'' (type II LAYM) theory the scalars are propagating. Also intermediary types exist 
\cite{Mayer-Strobl}.  Lie algebroids contain information of ordinary, finite dimensional Lie algebras living over the orbits in $M$. Thus, LAYM theories may glue together different YM-theories effectively \cite{LAYM}. This spurs hope to find new scenarios of WIMPs for dark matter.

The present considerations remained on the classical level. It will be important to go one step further and to study the behavior of the CYMH-theory on the quantum level. Note that naive methods of power counting will not be sufficient for proving renormalizability (say for a subclass of the theories): Already any analytical redefinition of the form $A^a \mapsto M^a_b(X) A^b$ of higher order in $X$, applied to a theory known to be renormalizable, destroys power counting renormalizability---while not changing its renormalizability. \cite{Tamarkin} provides a step towards such a refined theory of renormalization. 

The $B$-field contribution to field strengths of $A^a$ is necessary for curved deformations. It can disappear with deformation parameters as in  \eqref{Glambda} or remain also for $R_\nabla = 0$: in that case, the condition \eqref{curvature} reduces to 
$\g$-invariance of $B$, $B\in \Omega^2(M,\g)^G$, where $\mathrm{Lie}(G)=\g$. Even without deformation, such terms, with higher derivatives respecting the symmetries, can arise in the process of renormalization of sigma models like \eqref{Higgs}. In higher gauge theories, part of $B$ can also become dynamical \cite{Gruetzmann-Strobl}.

In the present paper, we provided a simple example of a CYMH theory. However, one can show much more \cite{inprep}: CYMH gauge theories exist for \emph{any} Lie algebroid that integrates to a generalization of a compact group, i.e.~to a so-called proper Lie groupoid \cite{MatiasRui}, provided the latter one admits a multiplicative bi-connection.

Finally, the ``adapted equivalence principle'' employed in this paper can be extended easily to include fermionic matter, as is necessary for realistic models of Nature.

\newpage
\begin{acknowledgments}
We are grateful to A.~Alekseev, M.~de Hoyo, R.~Fernandes, C.~Laurent-Gengoux, C.~Mayer, B.~Pioline, J.~Plefka, H.~Samtleben, J.~Stasheff, M.~Staudacher,  M.~Vasiliev and A.~Weinstein for interesting and motivating discussions or remarks. T.S.~wants to warmly thank several institutes and colleagues for their interest and great hospitality: at Berkeley, Berlin, Bonn, Curitiba, Geneva, Hannover, Moscow, Munich, PennState, Prague, Rio, Santo Andr\'e, and Vienna. This work was supported by the following grants: NSF PHY-1307408, CAPES PVE BEX348147, Swiss NSF 141329, the NCCR SwissMAP, and the  ERC MODFLAT.
\end{acknowledgments}

\bibliography{bibtexPRL}
\end{document}